\documentclass[%
    preprint,
 amsmath,amssymb,
prd,
floatfix,
]{revtex4-2}

\usepackage[utf8]{inputenc}%
\usepackage[dvipsnames]{xcolor}
\usepackage{graphicx}%
\usepackage{booktabs} 
\usepackage{titlesec}
\usepackage{hyperref}
\usepackage{siunitx}
\newcommand{\ud}{\mathrm{d}}
\titleformat{\subsubsection}
  {\normalfont\itshape}    %
  {\thesubsubsection}      %
  {1em}                    %
  {}                       %
\usepackage{enumitem}

\usepackage{todonotes}

\newcommand{\dpd}{d_{\text{pre-drift}}}
\usepackage{hyperref}%
\hypersetup{
    colorlinks=true,
    linkcolor=blue,
    filecolor=magenta,      
    urlcolor=cyan,
    citecolor = cyan,
    pdfpagemode=FullScreen,
    }

\usepackage{cleveref}
\usepackage{float}

\begin{document}

\title{Radiation reaction measurements via single-shot energy-loss determination in high-intensity laser-electron collisions}
\newcommand{\HIJ}{Helmholtz Institute Jena, Fröbelstieg 3, 07743 Jena, Germany}
\newcommand{\GSI}{GSI Helmholtzzentrum für Schwerionenforschung GmbH, Planckstraße 1, 64291 Darmstadt, Germany}
\newcommand{\IOQ}{Institute of Optics and Quantum Electronics, Max-Wien-Platz 1, 07743 Jena, Germany}

\author{Philipp Sikorski}
\email{ph.sikorski@gmail.com}
\affiliation{\HIJ}
\affiliation{\GSI}
\affiliation{\IOQ}

\author{Daniel Seipt}
\email{d.seipt@hi-jena.gsi.de}
\affiliation{\HIJ}
\affiliation{\GSI}
\affiliation{\IOQ}

\date{\today}

\begin{abstract}
    Effects of radiation reaction are considered a key factor for lasers-plasma-interactions at ultra-high intensities, and gaining precise understanding of these phenomena will be crucial for discoveries of novel strong-field QED effects.
    In this paper we investigate an experimental geometry that allows a direct single-shot measurement of radiation reaction energy losses in high-intensity laser-beam collisions.
    This is made possible by simultaneous measurements of the pre-collision and post-collision electron beam spectra in a single
    shot by employing a dedicated 90-degree scattering geometry between the electron beam and the colliding high-power laser.
    We discuss the principal requirements and design constraints, e.g.~for the electron beam divergence and source size. We derive an analytic expression for the correlation between the final particle energy and the scattering angle.
    Numerical simulations are performed employing both classical and stochastic quantum radiation reaction models to predict the experimental outcome and demonstrate the feasibility of the scenario.
  \end{abstract}

\maketitle

\section{Introduction}

    It is well known that radiation reaction (RR) and strong-field quantum electrodynamics (QED) effects are becoming more important in high-intensity laser-plasma interactions as more-and-more multi-PW laser facilities are coming online worldwide \cite{Danson2019}. 
    There is not only interest from a fundamental physics point-of-view about the nature of radiation reaction \cite{Bulanov:PRE2011,Ilderton:PLB2013,Harvey:PRL2017,ekman_reduction_2021,ekman_reduction_2022,bild_radiation_2019,gratus_maxwelllorentz_2022,holtzapple_significance_2022,hsiang_non-markovian_2022,quin_effect_2025}, but RR effects are expected to have also practical implications by affecting the various mechanisms of laser plasma acceleration and radiation generation at high intensity \cite{Koga:PhysPlas2005,Tamburini:NJP2010,Thomas:PRX2012,gonoskov_anomalous_2014,golovanov_radiation_2022,Gong:SciRep2019}.
    Therefore, high-precision tests of (quantum) radiation reaction have become a crucial aspect of the research programmes for (petawatt) high-intensity laser facilities \cite{mckenna2016high,sarri2025inputeuropeanstrategyparticle}.
    For further details on the subject and additional literature we refer to the reviews \cite{DiPiazza:RevModPhys2012,gonoskov_charged_2022,fedotov_advances_2023,popruzhenko_dynamics_2023,blackburn2020}.

    In simple terms, radiation reaction refers to the effect that the emission of radiation acts back on the dynamics of the charged particle emitting the radiation. This typically results in an energy loss \cite{Vranic_2016,BULANOV2024100036} or angular deflection \cite{Heinzl:2013,sikorski_novel_2024} of the emitting particles. A radiation dominated regime is reached when the radiative losses per laser cycle approach the initial particle energy, which occurs if the radiation reaction parameter $R_c \simeq \alpha a_0^2 \gamma \omega_0/m$ approaches unity \cite{Koga:PhysPlas2005,DiPiazza:LMP2008}, where $a_0$ and $\omega_0$ are the normalized laser potential and laser frequency, and $\gamma$ is the electron Lorentz factor, while $m$ is the electron mass; $\alpha$ is the fine structure constant.
    
    Quantum radiation reaction on charged particle electron dynamics is understood as the consecutive stochastic emission of photons \cite{DiPiazza:PRL2010}, resulting in stochastic energy loss and angular deflections due to the electron recoil experienced as a consequence of momentum conservation during photon emission. The quantum description becomes necessary when the parameter $\chi \simeq \gamma a_0 \omega/m$ approaches unity. The signatures of quantum effects have been studied, e.g. in \cite{Neitz:PRL2013,Blackburn:PRL2014,Ridgers:JPP2017,torgrimsson_quantum_2024}, and a detailed investigation of transition from the quantum to the classical regime has been presented in \citet{Niel}. In the classical limit, the effect of radiation reaction can be described as additional `friction' force-term in the classical equations of motion. As a concrete example where this distinction is significant, it should be noted that classical RR leads to a narrowing of electron energy distribution (beam cooling) \cite{yoffe_longitudinal_2015,bilbao_cooling_2023} while the stochastic nature of quantum RR causes a broadening of energy distribution (beam heating) \cite{Niel,blackburn2024}.

    Several experiments have been performed in the recent years with the aim of experimentally observing radiation reaction effects in high-power laser-particle interactions \cite{Cole:PRX2018,Poder:PRX2018,mirzaie_all-optical_2024}. Emphasis was placed into reaching the quantum regime $\chi\sim 1$, and distinguishing electron energy-loss signatures of classical and quantum radiation reaction regimes. However, the determination of the energy loss required not only to correlate the final electron spectra with the specta of emitted photons, but also to make assumption about shot-to-shot fluctuations of the incident electron beam. Elaborate statistical analysis including extensive simulation work was required for drawing quantitative conclusions \cite{los_observation_2026}.

    In this paper we investigate novel sceneario for high-precision experimental investigation of radiation reaction effects. It allows for a concurrent measurement of the electron beam energy before and after the interaction  and therefore a precise single-shot determination of the energy loss. Moreover, it is possible to straightforwardly post-select only those electrons for the analysis that have seen the highest laser intensity at the peak of the focus. We numerically simulated the scattering to produce virtual diagnostics demonstrating the capabilities of the method.

    The paper is organized as follows. In Section~\ref{sect:sidescattering} we describe the side-scattering scenario including design considerations and analytical estimates for the experimental outcome. Then in Section~\ref{sect:numerics} we present the numerical simulations before concluding in Section~\ref{sect:discussion}.

    \section{Single-shot radiation reaction energy-loss determination in side-scattering geometry}
    \label{sect:sidescattering}

    In typical radiation reaction (RR) experiments with high-power lasers, an electron beam---e.g.~from a laser wakefield accelerator---is collided (head-on) with a second scattering laser where the RR interaction is taking place. Then the post-scattering spectrum of the electrons that experienced RR is measured. In the `standard' approach, the precise determination of the electron energy loss can be cumbersome and require extensive statistical modeling \cite{Poder:PRX2018,Cole:PRX2018,los_observation_2026} since the initial pre-scattering spectrum is not exactly known for the very shot where the RR scattering took place; it has to inferred from other shots taking into account typical shot-to-shot fluctuations.

    Despite the tremendous progress achieved meanwhile in stabilizing LWFA, see e.g.~\cite{maier_decoding_2020,jalas_tuning_2023}, having a single-shot method available remains beneficial as it simplifies the data postprocessing and analysis considerably. To achieve this, it was proposed to measure the pre-scattering and post-scattering spectra \emph{in a single shot} \cite{Baird_2019}. This is achieved by first letting the electron beam expand due to its natural divergence. The scattering laser then only interacts with the central part of the electron bunch while the outer electrons remain unaffected by RR. This allows the simultaneous measurement of the initial and final energies for a straightforward determination of the RR energy loss. Here we develop further the ideas of \cite{Baird_2019} by going to a 90 degree scattering angle between the electron beam and the high-intensity scattering laser for the RR interaction, see Figure~\ref{fig:sckratch}. This allows to mitigate several of the challenges encountered in the head-on geometry of \citet{Baird_2019}.

    \subsection{The 90 degree scattering geometry}

    \begin{figure}[htb]
    \centering
    \includegraphics[width=1\linewidth]{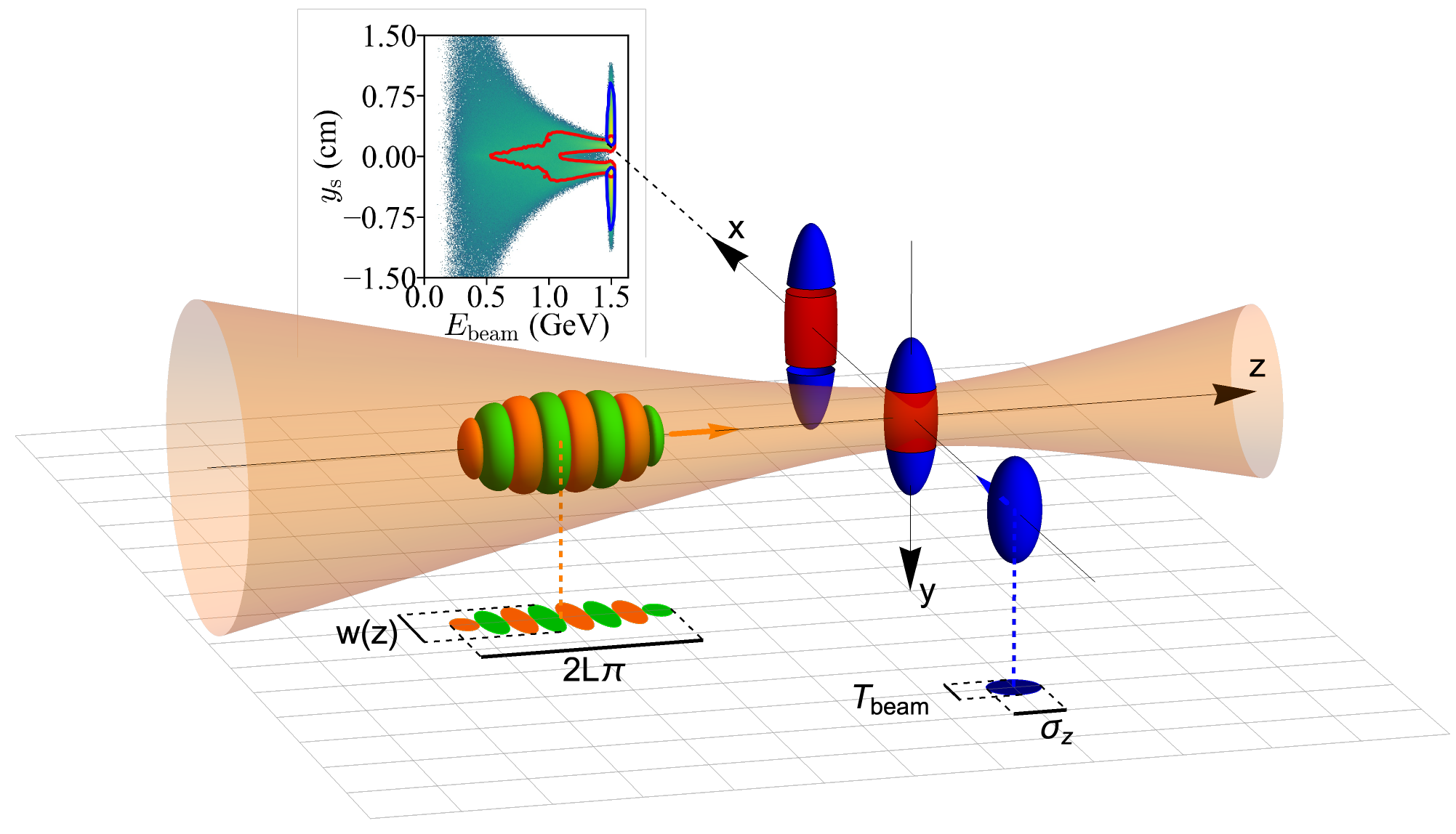}
    \caption{Layout of the 90 degree geometry for single-shot radiation reaction energy loss determination. The electron beam (blue) is allowed to expand before colliding at 90 degrees with the scattering laser pulse (orange/bright green). Only the central part of the expanded bunch interacts with the laser (red) allowing to measure simultaneously the pre-scattering and post-scattering spectra (see inset, red and blue contour lines).}        
    \label{fig:sckratch}
    \end{figure}

    The basic layout of the 90 degree single-shot RR energy loss experiment is exhibited in Figure~\ref{fig:sckratch}. LWFA electron beams typically have a micron-sized source size at exit of the plasma accelerator \cite{salgado_all-optical_2024}, small emittance, durations of sub-10 fs \cite{PhysRevSTAB.18.121302,seidel2025determinationfemtosecondattosecondelectron} and energy spreads on the few-percent level. If the beams are left free drifting the beam expands spatially and the transverse phase space becomes strongly correlated between $y$ and $\theta_y$ (cf. also Fig.~\ref{fig:phase-space}). 
    
    The three main criteria for the electron beam at the RR interaction laser can be summarized as follows: (i) its vertical size must be larger than the laser focal spot so that sufficiently many electrons miss the laser (ii) the beam must be sufficiently \emph{laminar}, i.e. the transverse phase space ellipse must be very thin (the local divergence much smaller than the overall beam divergence). (iii) It must be ensured that the inner (red) electrons don't mix `too much' with the outer electrons despite being ponderomotively scattered into the outside regions (blue regions in \Cref{fig:sckratch}. (See also Fig.~\ref{fig:phase-space}, the structures in the inner part are the scattered electrons distorting the ellipse.) We are quantifying these criteria below.

    When observed later on a screen, the uninteracted electrons still carry the initial energy spectrum, and they appear in the outer regions for large $|y_s|$. The particles that \emph{have} interacted with the scattering laser hit the detection screen at small $|y_s|$. From this inner region the post-collision spectrum can be inferred. Together, the initial and post-collision spectrum from the same shot allow for an accurate determination of the RR electron loss.

    \subsection{Electron beam and collision requirements}
    \label{sect:matrix}

    We consider the end of the LWFA stage as the source point (subscript $0$) for the electron beam with uncorrelated trace space $(y,y'=\theta_y)$, characterized by the covariance matrix
    \begin{align}
        \Sigma_0 
        = \left(
        \begin{matrix}
            \sigma_0^2 & 0 \\
            0 & \sigma_{0'}^2
        \end{matrix}
        \right) \,.
    \end{align}
    Here and throughout we employ the small angle approximation $\theta_y\approx p_y/p_x\ll 1$.
    The trace space density in one transverse dimension is given by
    \begin{align}
    n_0(y,y') = \exp \left\{ -\frac{1}{2} (y,y') \Sigma_0^{-1} (y,y')^T \right\} = e^{-\frac{y^2}{2\sigma_0^2} - \frac{y'^2}{2\sigma_{0'}^2} }\,.
    \end{align}
    
    The free drift of length $d_\text{pre-drift}$ between the source and the plane of the laser interaction (subscript 1) is described by the transfer matrix
    \begin{align}
        M_D = \left( 
        \begin{matrix}
            1 & d_{\text{pre-drift}} \\ 0 & 1 
        \end{matrix}\right) \,,
    \end{align}
    with $(y,y')^T \to  M_D (y,y')^T = (y + d_{\text{pre-drift}} y', y')^T = (y + d_{\text{pre-drift}} \theta_y, \theta_y)^T$.
    Therefore, at the scattering laser the beam matrix is
        \begin{align} 
        \Sigma_1 = M_D \Sigma_0 M_D^{T}  =
        \left( 
        \begin{matrix}
         \sigma_0^2 +  d_{\text{pre-drift}}^2 \sigma_{0'}^2 & d_{\text{pre-drift}} \sigma_{0'}^2 \\
         d_{\text{pre-drift}} \sigma_{0'}^2 & \sigma_{0'}^2
        \end{matrix}\right)\,,
    \end{align}%
    with trace space density
    \begin{align} \label{eq:n1}
        n_1(y,y') = e^{-\frac{(y - d_{\text{pre-drift}} y' )^2}{2\sigma_0^2} - \frac{y'^2}{2\sigma_{0'}^2} }
    \end{align}
    is a tilted ellipse with a strong correlation between the the electron location and angle. The geometric rms emittance $\epsilon=\sqrt{\text{det} \Sigma }=\sigma_0\sigma_0'$ remains constant.

    At the laser scattering plane the beam is characterized by the Courant–Snyder parameters
    \begin{align}
        \alpha &= - \frac{\dpd \sigma_{0'}}{\sigma_0} \,, \\
        \beta &=  \frac{\sigma_0}{\sigma_{0'}} + \frac{\dpd^2 \sigma_{0'}}{\sigma_0} \,, \\
        \gamma &= \frac{\sigma_{0'}}{\sigma_0}\,.
    \end{align}
    The beam size has expanded to spatial size $\sigma_1 =  \sqrt{\epsilon\beta} = \sqrt{\sigma_0^2 + d_{\text{pre-drift}}^2\sigma_{0'}^2}$. The local angular spread is $\sigma_{1'}(y) = \sqrt{\epsilon/\beta} = (d_{\text{pre-drift}}^2/\sigma_0^2 + 1/\sigma_{0'}^2)^{-1/2}$, much smaller than $\sigma_{0'}$; the beam expands in a `laminar' fashion at the laser interaction plane, while the overall rms divergence has not changed, $\sigma_{1'}=\sigma_{0'}$. This correlation is crucial; the local angular spread must be small compared to the total angular spread, i.e.~the covariance ellipse must be `very thin`---the $\beta$ function must be large.
        
    The drift distance is chosen such that the beam size in the $y$-direction at the laser interaction point $\sigma_{y,1} > w_0$, where $w_0$ is the laser focal spot only the central part of the electron beam (colored in red in Fig.~\ref{fig:sckratch}) will interact with the laser and experience energy loss due to radiation reaction. The electrons in the wings of the beam will miss the laser and retain their pre-scattering spectra. If the subsequent electron spectrometer measurement disperses the electron energy perpendicular to the $y$-axis both the pre-scattering and post-scattering spectra could be measured simultaneously without much blurring between the two fractions. In the $z$-direction the requirements for the beam size are less crucial, the main criterion being that the spatial size is smaller than the laser Rayleigh range, $\sigma_{z,1} < z_R$.

    During the interaction with the laser, the electrons will also be angularly deflected, mainly due to the ponderomotive force (see also Section \ref{sect:analytics}). This increases their angles and causes a `leakage' into the regions of uninteracted electrons during the post-interaction drift to the detection screen. If that happens, it will not be possible to distinguish the pre-interaction and post-interaction spectra uniquely.

\begin{figure}[H]
    \centering
    \includegraphics[width=0.7\linewidth]{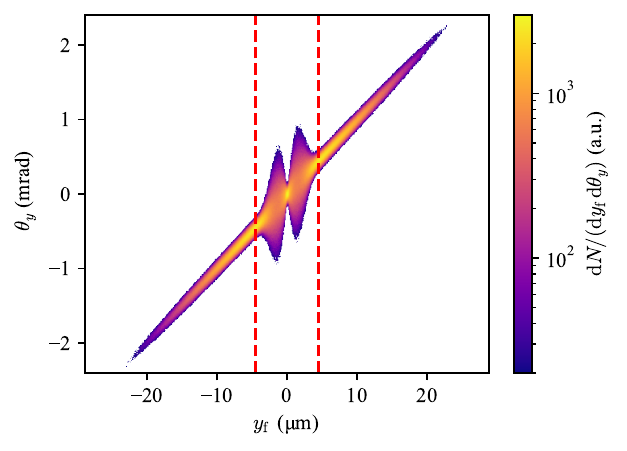}
    \caption{Transverse electron beam phase space directly after the interaction at the end of the PIC simulation box. Vertical dashed lines mark the location of the scattering laser focus strongly impacting the beam. In the central part ponderomotive scattering deflects electrons changing their $\theta_y$. The scattering geometry has to be chosen such that the post-interaction drift does not blur them into the spatial regions of uninteracted electrons. }
    \label{fig:phase-space}
    \end{figure}

    {If we consider $\sigma_{y,1} > w_0$ and the beam emittance as fixed parameters, then this criterion dictates that the inital beam divergence $\sigma_{y,0'}$ has to be larger than the angular deflection due to ponderomotive scattering (see \cref{fig:phase-space}). That means if we keep the beam emittance fixed then the source size $\sigma_{y,0}$ should be very small,  $\sim \qty{1}{\micro m}$, as is typical for LWFA beams~\cite{salgado_all-optical_2024,salgado_limitations_2026}.}

       \subsection{Advantages of the 90 degree scattering geometry}

    Despite being conceptually quite similar to the concept put forward in Ref.~\citet{Baird_2019}, the 90 degree scattering geometry brings a number of benefits compared to the head-on collision.
    First, due to the 90 degree geometry the electron beam does not go along the laser axis after the interaction as it would be the case for the head-on scattering. This completely eliminates the need for laser optics in the electron beam path that could be the source of additional scattering.

    Second, the relative timing jitter between the electron and laser beams could be an issue in both scenarios, but has very different outcomes. In the \citet{Baird_2019} case, a temporal offset would mean that the electrons interact with the laser pulse before or after it arrives at its focus location, i.e. the interaction occurs not at the full laser intensity. 
    Contrary, in our 90 degree geometry it would mean that the beams would miss each other completely if the laser pulse duration is shorter than the timing jitter. This can be easily mitigated by using longer pulse durations for the scattering laser (e.g. $\ge100$ fs). Moreover, at 90 degrees the electron beam pointing jitter is relevant strongly only along the $y$-axis, perpendicular to the laser beam axis, while for head-on one has to worry about both transverse directions.

    In the experiment, the particle spectra cannot be measured directly after the interaction, the beam has to go through a magnetic spectrometer first where they are dispersed along a certain direction. In the head-on geometry, the energy dispersion causes a mixing/blurring of interacted and non-interacted electrons on the detection screen due to the cylindrical symmetry of the ring-shaped region of un-interacted electrons around the laser focus. Contrary, in our 90-degree geometry this can be completely mitigated due to the linear stacking of interacted and non-interacted electrons along the $y$-axis by aligning the energy dispersion axis perpendicular to the $y$-axis.
    
    Finally, it should be feasible to post-select only those electrons that have interacted with the highest intensity $a_0$ at the peak of the focus. This is achieved by selecting only the small band around $\theta_\text{s}=0$ on the detection screen as we do in the analysis further below in Section~\ref{sect:numerics}. This means we do not need to perform deconvolution on the measured electron energy distributions, but instead obtain a direct result. In head-on collision scheme proposed by \cite{Baird_2019} this is not simlpy possible as energy spectrometer disperses electrons interacted with variable effective $a(r)$ onto the same location on the detection screen due to the cylindrical symmetry.

    We also want to mention the drawback of going away from head-on collisions: The quantum parameter for an electron-laser collision angle $\Theta$ is smaller than head-on, namely 
    \begin{align}
    \frac{\chi (\Theta)}{\chi_\text{head-on}} \simeq 
    \cos^2 \frac{\Theta}{2}\,.
    \end{align}

    \subsection{Analytical estimate for the expected correlation between  energy-loss and deflection angle}
    \label{sect:analytics}
    
     Here we derive an analytical model to qualitatively describe the nontrivial shapes of the final electron distribution in $y_\text{s}$ and $E_{\text{beam}}$ as they emerge from an interplay between radiation reaction effects and ponderomotive deflection (numerical results are shown below in \Cref{fig:2x5})
     We employ the semi-classical Landau-Lifshitz equation, which accounts for the reduction of radiated power due to quantum effects,
     to model the effects of radiation reaction.

    The nontrivial shapes of the distributions arise because the electron energy changes mainly due to radiation reaction (RR) effects, while $y_\text{s}$ is primarily influenced by the laser ponderomotive force. These two effects scale differently with the laser intensity. In this section, we aim to predict the final $y_\text{s}$-position and the energy of a particle with initial position $y_1$ (and assumed to be located at the horizontal center of the focus when the temporal peak arrives).

    An electron arriving at the plane of the scattering laser at vertical location $y_1$ will experience an effective peak normalized vector potential determined by the Gaussian laser focus profile, according to
    \begin{align}
    \label{eq_a_y}
    a_1 \equiv 
    a_0 e^{-y_1^2 / w_0^2}\,.
    \end{align}

    Treating the laser locally as a plane wave we can employ the known analytic solution of the Landau-Lifshitz equation from the literature \cite{DiPiazza:LMP2008} with the the local peak value $a_1$. This yields the final electron Lorentz factor as
    \begin{equation}
    \label{eq_gamma_approx} 
    \gamma \simeq \frac{\gamma_0}{1 + \frac{2}{3} R_c g(\bar \chi_1) \mathcal {I}}\,,
    \end{equation}
    where $R_c = \alpha a_1^2 \omega_0 \gamma_0 / m$ is the classical radiation parameter and
    \begin{align}
    \label{eq_gaunt} 
    g(\chi) = \left[1 + 4.8 (1 + \chi) \log(1 + 1.7 \chi) + 2.44 \chi^2\right]^{-2/3}
    \end{align}
    is the Gaunt-factor \cite{gonoskov_charged_2022}, with the effective time-averaged quantum nonlinearity parameter given as $\bar \chi_1 = a_1\gamma_0 \omega_0/(m{\sqrt{2}})$.
    In \cref{eq_gamma_approx} we also defined the normalized integrated laser intensity experienced by the particle as it traverses the focus
     $\mathcal{I} = \omega_0^{-1} \int_{-\infty}^{\infty} [\ud f(t,x,0)/\ud t]^2|_{x=t}\,\mathrm{d}t$ at $z=0$ on a trajectory $x(t)\approx t$, with
    \begin{equation}    
    f(t,x,z)=\cos^4\left(\frac{\omega_0 (t-z) } {2 L}\right) \: \cos\omega_0 (t-z) \: e^{-\frac{x^2}{w_0^2}}
    \end{equation} 
    being an approximate expression for laser vector potential shape function in the vicinity of the focus for $z\ll z_R$. The dependency on $y_1$ has been absorbed into the definition of $a_1$, and in this model we assume that $y_1=const.$ during the interaction with the laser.
    
    The final vertical electron momentum $u_y$ is dominated by the ponderomotive force $-\frac{\nabla a(x,y_1)^2}{4\gamma}$, and it's value is here approximated as 
    \begin{align}
     \label{eq:uy_ps_rr}
        u_y(y_1) &= 
        u_{y,0}(y_1)+\frac{y_1}{w_0  } \frac{a_1^2}{\bar \gamma}\sqrt{\frac{\pi}{2}}\,,
    \end{align}
    where $u_{y,0}$ is the initial electron momentum prior to the laser scattering and $\bar \gamma = (\gamma+\gamma_0)/2$ is the mean energy. From this, by taking into account that $u_x=\sqrt{\gamma^2-1-u_y^2}$ we can find the final vertical electron position on the detection screen given by
    \begin{equation}
    \label{eq:ys}
        y_\text{s} = y_1 + d_\text{post-drift}\: \frac{u_y(y_1)}{u_x(y_1)}\,.
    \end{equation}

\begin{figure}[H]
    \centering
    \includegraphics[width=0.7\linewidth]{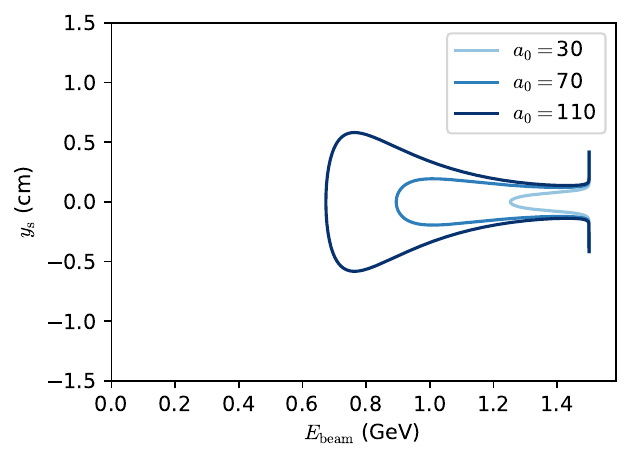}
    \caption{The shape of the angular-energy distribution obtained with \cref{eq_gamma_approx} and \cref{eq:ys}.}
    \label{fig:explanation}
    \end{figure}

    By using \cref{eq_gamma_approx} and \cref{eq:ys} we obtain the correlation of final $E_\text{beam}=m\gamma$ and $y_\text{s}$ resembling characteristic droplet shapes.
    They are exhibited for various values of $a_0$
    in \cref{fig:explanation}. For lower values of $a_0$ the droplets taper to a point for small $E_\text{beam}$, while for 
    larger $a_0$ the droplets widen significantly at lower $E_\text{beam}$. The corresponding distributions obtained via numerical simulations look very similar (see \cref{fig:2x5} below in \cref{sect:numerics}).

    \section{Numerical results and discussion}

    \subsection{3D QED simulations with SMILEI}
    \label{sect:numerics}

    We performed 3D Monte-Carlo simulations of the electron-beam laser interactions using the Particle-in-Cell code SMILEI \cite{smilei}. The simulation box size was $l_x=28.8\,\si{\micro m}$, $l_y=57.6\,\si{\micro m}$, $l_z=14.4\,\si{\micro m}$ with a resolution of $\Delta x = 400\,\si{\nano m}$, $\Delta y = 400\,\si{\nano m}$, $\Delta z = 25\,\si{\nano m}$ and $\Delta t \approx 82\,\si{\atto s}$.
    The laser pulse propagates along the $z$-axis, has a central wavelength $\lambda=800\,\si{\nano m}$ and a beam waist of $w_0=2.5\,\lambda=2\,\si{\micro m}$. It is linearly polarized along the $x$-direction, with a normalized vector potential varied in the range $a_0=30\ldots 110$. The pulse temporal envelope was chosen as $\cos^4$-shaped containing a total of 100 optical cycles. This  corresponding to a full width at half maximum (FWHM) duration of $\tau_{\mathrm{FWHM}} \approx 98\,\si{\femto\second}$.

    The electron beam propagates along the $x$-axis with the initial energy of $E_0=\qty{1.5}{\giga eV}$ ($\gamma_0 = 2935$) and an energy spread of $\Delta E=\qty{8}{\mega \electronvolt}$. It is modelled as a Gaussian beam with transverse sizes $\sigma_{z,1} = 1\,\si{\micro m},\ \sigma_{y,1} = 7\,\si{\micro m}$ and the duration $T_{\mathrm{beam}} = 2\,\si{\femto s}$ at the interaction point with the scattering laser. Hence, the criterion $\sigma_{y,1} > w_0$ is ensured. 
    The expanding electron beam is initialized in the simulation as follows: We define the rms beam size at the laser focus plane, $\sigma_{x,1},\,\sigma_{y,1},\,\sigma_{z,1}$ and also at the accelerator exit, $\sigma_{x,0},\,\sigma_{y,0},\,\sigma_{z,0}$, located at a distance $d_{\text{pre-drift}}=\qty{10}{\milli m}$ in front of the simulation box. The transverse phase spaces are assumed to be uncorrelated Gaussians at the accelerator exit. We then calculate the corresponding momentum distributions of the electron beam required to accommodate those spatial sizes in both planes ``0'' and ``1'' in each transverse direction, see \Cref{sect:matrix}:
    \begin{align}
      \sigma_{p,0} = \gamma_0 \sigma_{0'} = \gamma_0 \frac{\sqrt{\sigma_1^2 - \sigma_0^2}}{d_{\text{pre-drift}}} \,.
    \end{align}
    The transverse electron momenta, normalized to $mc$, are sampled from Gaussian distributions with standard deviation $\sigma_{p,0}$.
    In this way, by changing the size $\sigma_{0}$ at the accelerator exit we control the divergence of the beam. 
    The normalized longitudinal momentum is defined according to $p_x=(\gamma_0^2-1-p_y^2-p_z^2)^{1/2}$ in order to keep the total energy constant.
    For the simulations presented below we utilize $\sigma_{y,0} = \qty{0.5}{\micro m}$, hence $\sigma_{y,0'}=\qty{0.69}{\milli\radian}$ and $\sigma_{z,0} = \qty{0.5}{\micro m}$ with $\sigma_{z,0'}=87\,\si{\micro\radian}$. A more detailed discussion of the relevance of the initial beam divergence is presented in \Cref{app:a}.

    The radiation reaction is simulated using two different radiation models: the semiclassical (quantum-corrected) Landau-Lifshitz [SCRR] and the quantum Monte–Carlo (quantum stochastic) [QRR]. The scattered electron beam at the end of the PIC simulation is then propagated towards a detection screen placed at a distance $d_{\text{post-drift}}=\qty{4}{m}$ behind the laser focus. At the screen location we investigate the particle distributions in energy vs. the {$y$-coordinate} on the detection screen $y_\text{s}$. Note that we do not explicitly simulate the electron spectrometer; the particle energy distributions are taken directly from the PIC simulation output. The $y_\text{s}$-coordinate is calculated using the following eqution:
    \begin{equation}
        y_\text{s}=\theta_y d_{\text{post-drift}}+y_\text{f}\,,
    \end{equation}
    where $y_\text{f}$ is the $y$-coordinate and 
    $\theta_y = u_y/u_x$ is the vertical electron angle, both taken at the end of the simulation.

    \subsection{Simulation results}

\begin{figure}[!thb]
    \centering
    \includegraphics[width=1\linewidth]{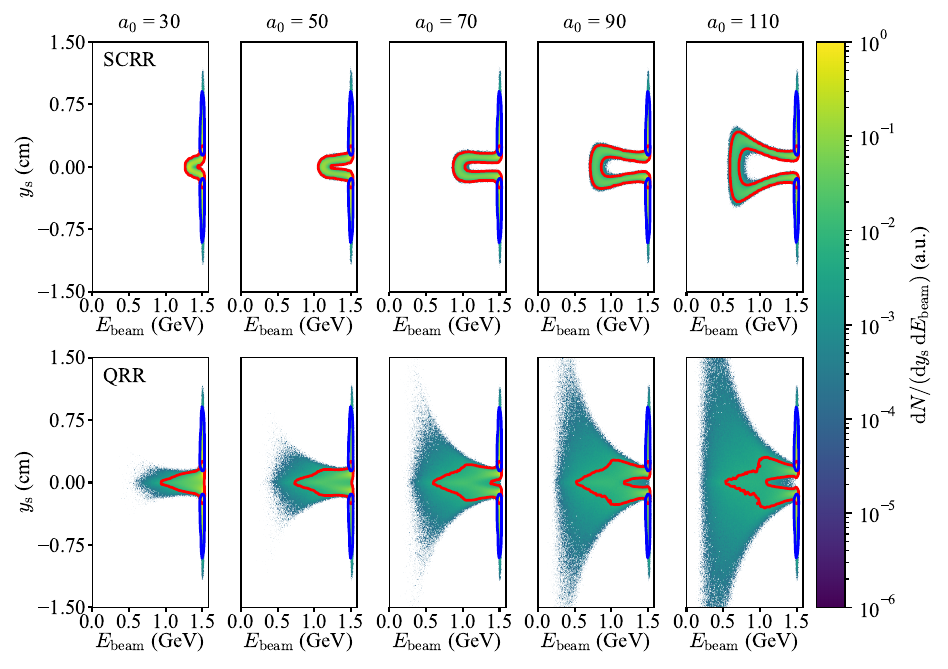}
    \caption{Particle distributions $y_s$ vs. $E_\text{beam}$ observed on a detection screen 4 meters behind the interaction point for various values of $a_0$. Top row is for semiclassical radiation reaction (SCRR) and the bottom row exhibits the results for stochastic quantum radiation reaction (QRR). The blue and red curves indicate the regions where unscattered and scattered electrons dominate (compare with Figure~\ref{fig:sckratch}).
    }
    \label{fig:2x5}
    \end{figure}

    The numerical results for the $y_s$ vs. $E_\text{beam}$ distributions on the detector screen are exhibited in \Cref{fig:2x5} for increasing values of $a_0$ from left to right. The SCRR model is presented in the upper row and for the stochastic QRR model is shown in the lower row. In each case the electron distributions show characteristic patterns that allow a distinction of the two electron fractions that have interacted with the scattering laser and those who have not.
    
    The latter uninteracted-electron-feature appears as two relatively narrow vertical lines at the primary energy $E_0=\qty{1.5}{GeV}$ in all panels at $|y_s| \gtrsim \qty{0.25}{cm}$. Such features should allow a clear determination of the primary electron energy on a shot-to-shot basis.

    The features of the scattered electrons change depending on the radiation reaction model, and the specific value of $a_0$. For SCRR the interacted electrons there is a strong correlation between $y_s$ and $E_\text{beam}$. The distribution follows a characteristic droplet shape, which looks similar to the shapes derived above in \ref{sect:analytics}. The main difference is an additional spreading in $E_\text{beam}$ and $y_\text{s}$ due to the initial momentum distributions.

    For QRR (lower panels) the stochastic nature of photon emission changes the picture. The electron distributions no longer follow the analytic curves, but rather are spread out over a whole area shaped similar to these curves. Some of the electrons have lost more energy and some are scattered to much larger $y_s$ than in the SCRR case. Nonetheless, the features from the pre-interaction spectra are clearly distinct from the scattered electron features.

    \subsection{Intensity post-selection}

\begin{figure}[H]
        \centering
            \includegraphics[width=1\linewidth]{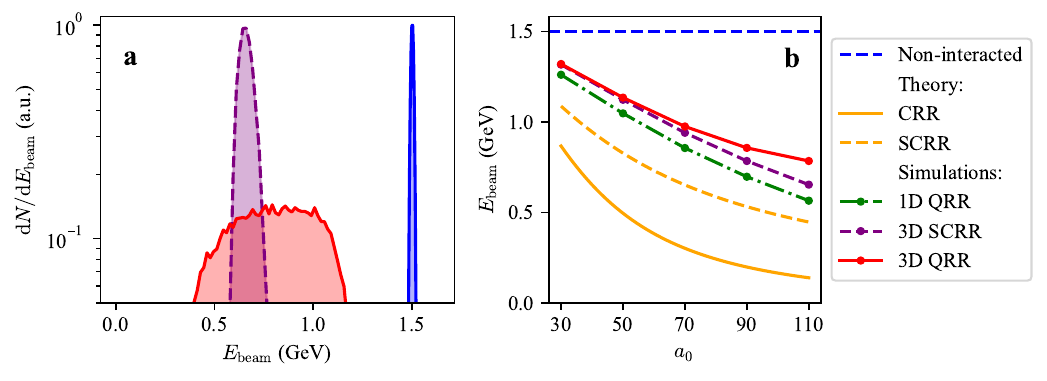}
        \caption{
        Left panel: Post-selected spectra of interacted electrons from the SCRR and QRR simulations (purple and red curves), respectively, and the pre-interaction spectrum extracted from the non-interacted electrons (blue curve) for the case of $a_0=110$, corresponding to the rightmost column in \Cref{fig:2x5}.
        For the interacted electrons, the spectrum is obtained by postselecting for those particles with transverse screen coordinates $y_s$ within the range $[\qty{-100}{\micro m}, \qty{100}{\micro m}]$. 
        Right panel: Predictions of the final beam energy for different models as a function of $a_0$. For the simulation results, the mean energy is calculated from the spectrum of particles with transverse screen coordinates $y_s$ within the range $[\qty{-100}{\micro m}, \qty{100}{\micro m}]$.  For the theoretical results, the energy loss is calculated for a single test particle.
        }
        \label{fig:spectra}
    \end{figure}

   With the 90 degree scattering scenario we can also post-select for electrons that have centrally-hit the laser focus and therefore interacted with the peak value of $a_0$, and did not scatter to large angles. In this way we may achieve a partial deconvolution of the laser focus intensity distribution. We can take, for instance, only those electrons with $|y_s| < \qty{100}{\micro m}$, that have seen the highest laser intensities near the focus, and determine their spectrum. This is shown in \Cref{fig:spectra} a) where we compare the pre-collision spectrum (blue curve) with the post-interaction spectra according to the SCRR and QRR models (purple and red curves, respectively).
    
    We see that the SCRR peak is much narrower than the QRR peak, and the mean energy of the SCRR peak is lower than the QRR peak. This is in agreement with theoretical expectations (see e.g.~\cite{Ridgers:JPP2017}). However, we see that the SCRR peak is in fact wider than the pre-interaction spectrum. According to theory predictions the SCRR radiative losses one would expect a narrower peak than the pre-interaction spectra due to classical beam cooling. However, due to the three dimensional interaction geometry, and a certain range of positions along the laser propagation axis being probed by the electrons, the resulting spectrum is comprised of a residual convolution of laser intensities near the focus. Consequently, the centrally-hit electrons spectrum is actually broader than the pre-interaction one.

    We exhibit the predictions for the radiative energy losses according to various models and simulations as a function of $a_0$ right panel of \cref{fig:spectra}.

    For CRR and SCRR, we calculated the energy loss using \cref{eq_gamma_approx}, derived above in \Cref{sect:analytics} as a solution of the Landau--Lifshitz equation. For CRR, the Gaunt factor was set to unity, whereas for SCRR, we used \cref{eq_gaunt}, taking
  $  \chi = a_1\gamma_0\omega_0/m \sqrt{2}$, see \Cref{sect:analytics}.
    All other parameters entering \cref{eq_gamma_approx} were calculated following the same procedure as in \cref{sect:analytics}, considering only one particle centrally hitting the laser spot on axis.
    As expected, the analytical predictions yield the largest energy losses, since they correspond to an idealized case of a single particle experiencing the peak laser intensity throughout the interaction. SCRR, however, shows much better agreement with the simulation results, as it accounts for the quantum suppression of radiation losses through the Gaunt-factor correction. 
    The curve `1D QRR' corresponds to the results of 1D PIC simulations, in which all particles have identical constant $x$ and $y$ coordinates and non-zero $p_x$ momentum. This model accounts for the stochastic nature of quantum radiation emission, but neglects three-dimensional geometrical effects of the collision. 
    The 3D simulations labeled SCRR and QRR correspond to the results of the 3D simulations presented in \Cref{fig:2x5}. As can be seen, especially at smaller $a_0$, 
    the curves from the SCRR and QRR 3D simulations agree very well, indicating that the effect of stochasticity is small in this region.
    Contrary, the influence of the 3D geometry (SCRR theory vs. 3D SCRR simulations) remains approximately constant over the whole parameter region. The relevant 3D effects, include the fact that different particles experience different laser intensities and interaction times, rather than all particles remaining at the peak intensity for the entire duration of the interaction.

    \section{Summary and Conclusions}
    \label{sect:discussion}

    In this article we put forward a novel scheme for a single-shot determination of the radiative electron loss in electron-laser collisions by measuring simultaneously the pre-collision and post-collision spectra. Our method uses a 90 degree scattering configuration between an LWFA electron beam and intense laser pulse. The electron beam is pre-expaned before interacting with the RR laser, such that part of the beam is passing the focal plane unaffected and still carry the pre-collision spectra to be measured. Such a collision requires sufficiently large initial beam divergence and small source size.

    We performed 3D Monte Carlo simulations of the RR interaction for a semiclassical and quantum stochastic model of radiation reaction over a wide range of laser intensities. Each of them generates characteristic correlations between the electron energy and scattering angle. For the semi-classical RR case 
    we analytically derived the characteristic droplet-like shape using a simple model. In all considered cases there are distinct features of the pre-collision spectra visible in on the synthetic detection screens. Thus, our numerical results demonstrate that such simultaneous detection of the pre-collision and post-collision spectra simultaneously is possible. Together with a post-selection of centrally-hit electrons a clear distinction between the different radiation reaction regimes should be possible.

    Realizing such an experiment at a multi-beam high power laser facility seams feasible in the near future (e.g.~ELI, Apollon). The set-up is relatively insensitive to spatial jitter between the electrons and the laser, as this would just mean the laser hits on one side of the electron beam. The collision and corresponding particle distributions on the detection screen become asymmetric, but the pre-spectra can still be distinguished from the post-collision one. Moreover, the experimental set-up with the 90 degree collision angle makes the experiment quite insensitive to the timing jitter between the electron beam and the scattering laser if long ($\gtrsim\qty{100}{fs}$) pulses are employed for the scattering laser.

    \begin{acknowledgements}
    The authors thank Aimé Mathreron for useful fruitful discussions and comments on the manuscript. 
    The authors gratefully acknowledge the Gauss Centre for Supercomputing e.V. (www.gauss-centre.eu) for funding this project by providing computing time through the John von Neumann Institute for Computing (NIC) on the GCS Supercomputer JUWELS at Jülich Supercomputing Centre (JSC). The research leading to the presented results received additional funding from the European Regional Development Fund and the State of Thuringia (Contract No. 2019 FGI 0013).
    \end{acknowledgements}

    \appendix

    \section{Importance of the initial beam divergency}

    \label{app:a}

    From the phase space shown in \Cref{fig:phase-space}, we can see that particles that have interacted with the laser exhibit a much larger divergence than they had initially. This has also been quantified by the analytic model, especially \Cref{eq:uy_ps_rr} in \Cref{sect:analytics}. 

    This deflection of the particles could cause some issues, as these scattered particles could leak into the outer regions of the detector 
    where we want to see only particles that do not interact with the scattering laser. But it is also especially problematic when analyzing the spectrum of the small central part of the beam to select only those particles that have interacted with the highest laser intensity and therefore experienced the largest energy loss, see
    \Cref{fig:spectra}. If the initial beam divergence of is not sufficiently large, these particles will remain within the selected region and could reduce the observed energy loss.

\begin{figure}[H]
    \centering
    \includegraphics[width=\linewidth]{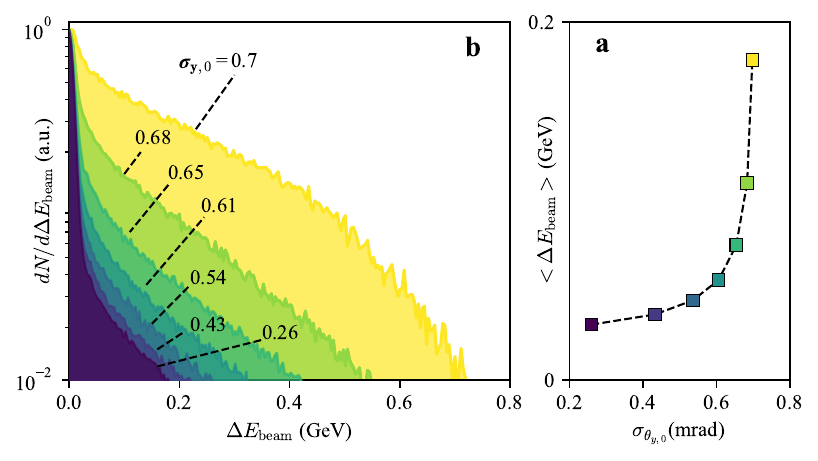}
    \caption{Observed mean energy loss of electrons collected over a range $\qty{\pm100}{\micro m}$ from the central axis for different initial electron beam divergences: (a) Energy loss spectrum for different initial divergences. (b) Mean value of the energy loss versus initial divergence.}
    \label{fig:initial_divergence}
    \end{figure}

    In \Cref{fig:initial_divergence}, we simulated several different initial divergences and obtain the corresponding spectra for the region $[\qty{-100}{\micro\meter}, \qty{100}{\micro\meter}]$. Note that all spectra are normalized. As we can see, by increasing the initial divergence, we increase the fraction of particles in the total energy-loss spectrum that have lost more energy due to radiation reaction. As a result, a larger initial divergence of the electron beam leads to a greater mean energy loss in the central part of the beam.
    
\bibliography{references}
\end{document}